# A Low-Voltage, Low-Power 4-bit BCD Adder, designed using the Clock Gated Power Gating, and the DVT Scheme

Dipankar Saha[1], Subhramita Basak[1], Sagar Mukherjee[2], C. K. Sarkar[1]
[1]Department of Electronics and Telecommunication Engineering
Jadavpur University
Kolkata, West Bengal, India
[2]Department of Electronics and Communication Engineering
MCKV Institute of Engineering
Liluah, Howrah, West Bengal, India
dipsah_etc@yahoo.co.in, basaksubhramita@gmail.com, sagarju87@gmail.com,
phyhod@yahoo.co.in

*Abstract*—This paper proposes a Low-Power, Energy Efficient 4-bit Binary Coded Decimal (BCD) adder design where the conventional 4-bit BCD adder has been modified with the Clock Gated Power Gating Technique. Moreover, the concept of DVT (Dual-$v_{th}$) scheme has been introduced while designing the full adder blocks to reduce the Leakage Power, as well as, to maintain the overall performance of the entire circuit. The reported architecture of 4-bit BCD adder is designed using 45 nm technology and it consumes 1.384 µWatt of Average Power while operating with a frequency of 200 MHz, and a Supply Voltage ($V_{dd}$) of 1 Volt. The results obtained from different simulation runs on SPICE, indicate the superiority of the proposed design compared to the conventional 4-bit BCD adder. Considering the product of Average Power and Delay, for the operating frequency of 200 MHz, a fair 47.41 % reduction compared to the conventional design has been achieved with this proposed scheme.

## I. INTRODUCTION

There are quite a few reasons behind the use of binary data for doing the arithmetic operations in almost all the computer systems. The speed and simplicity of binary arithmetic, efficiency in storing the binary data etc. are the most important among those [1, 2]. But, for the financial and commercial applications, the use of decimal arithmetic is still relevant. In case of the financial applications, in order to obtain the decimal results, a decimal software use to run on top of the underlying binary hardware. But the speed (which is almost 100 to 1000 times slower) is the major concern for the decimal softwares [3, 4]. Moreover, the commercial databases contain more decimal data than binary data. For the purpose of processing, these decimal data are converted into binary data. And, once the processing is completed, those are again converted back into the decimal format. The conversion between the decimal and the binary formats causes a significant amount of Delay [5]. Again, when the conversion takes place for the fractional decimal numbers such as 0.110, 0.1 etc., there occurs error (which is not tolerable for most of the financial and commercial applications) due to the approximated representations [5, 6]. Therefore, for all those reasons as mentioned above, designers have to look for a decimal arithmetic hardware in financial and commercial applications.

Now, in all arithmetic units, whether binary or decimal, the use of an adder circuit is obvious. So far, different techniques have been invented and proposed in literature for the purpose of doing the decimal additions (even though it is less popular than the binary addition) [5], [7-11].

The conventional 4-bit BCD adder consists of two 4-bit full adders and a carry detection logic circuit [9]. At the first stage, one of the two 4-bit full adders can be used to produce the binary addition results. If the value of the result obtained, is greater than the decimal number '9', then a carry output is generated. As well as, the result needs to be corrected by adding the decimal number '6' with it. This is actually done, at the next stage, by the other one of those two 4-bit full adders. Besides, a carry detection logic circuit is there, which is formed using two AND gates and one OR gate. In [10], Shirazi, et. al., proposed a Redundant Binary coded Decimal (RBCD) adder, where the addition operation is done in three steps. Firstly, the BCD input is converted into RBCD. Then, in the next stage, the result of the addition is obtained using the RBCD adder. At the final step, this RBCD adder-result is again converted back into the BCD format. But, as the conversion of RBCD adder-result back into the BCD format requires carry propagation, therefore the Delay caused due to this, becomes dependent on the length of the input operands [10]. In [11], a 4-bit BCD adder with Clock Gated Power Gating has been reported using 65 nm technology. As well as, a performance analysis, showing the comparison among "Conventional BCD adder", "BCD adder with DSTN (Distributed Sleep Transistor Network)", and "BCD adder with Clock Gated Power Gating" has been presented there. Now, considering the Average Power consumption and the Delay, as shown in [11], the 4-bit BCD adder with Clock Gated Power Gating is happened to be the most efficient one out of those three.

In our proposed architecture for the 4-bit BCD adder, we have mainly modified the conventional 4 bit BCD adder design [9] with the Clock Gated Power Gating. Besides, the entire design is carried out in 45 nm technology, and thereby a serious consideration about the means by which the Leakage current can be reduced becomes very much important. We introduced the concept of the DVT (Dual-$v_{th}$) scheme in our proposed design for the purpose of reduction of the Leakage Power. Moreover, to verify the efficiency of the proposed architecture, a comparison of this with the conventional 4-bit BCD adder and the 4-bit BCD adder with DVT scheme (without Power Gating) has also been presented.

Rest of the paper is organized as follows. In section II, a detailed discussion about the effectiveness of the DVT (Dual-$v_{th}$) scheme in reducing the Leakage Power, has been presented. In section III, a brief review of the several Power Gating strategies available in literature has been described. The details of the proposed design of 4-bit BCD adder have been illustrated in section IV. Section V, deals with the analysis of the results obtained from different simulation runs. Finally, in section VI, the conclusion summary of the entire work, has been discussed.

## II. LEAKAGE POWER & DVT SCHEME

For the digital circuits used in VLSI, power dissipation can be categorized mainly into three parts, Switching Power dissipation, Short-Circuit Power dissipation, and Static Power dissipation [12]. While considering the power dissipated by a MOS device, until recent years, the Static Leakage Power component has been assumed to be negligible. But, in the cases where we have to consider the sub-micron technologies, or, deep-sub-micron technologies, this assumption proving to be no longer true. Now, in standby mode, the power dissipation that occurs in a CMOS circuit is mainly due to the sub-threshold leakage current. The sub-threshold current of a MOSFET can be modeled as [13],

$$I_{sub} = A.e^{(q/n'KT)).(V_G - V_S - V_{To}'V_S + \eta V_{DS})} \left(1 - e^{-qV_{DS}/KT}\right) \quad (1)$$

Where, the drain to source voltage is represented as $V_{DS}$, and the gate voltage is denoted by $V_G$.

$$A = \mu_0 C_{ox} \frac{W_{eff}}{L_{eff}} \left(\frac{KT}{q}\right)^2 \times e^{1.8},$$

$C_{ox}$ is the per unit area gate oxide capacitance, $\mu_0$ is the zero bias mobility [13].

And as there exists an exponential relationship of the leakage current to the change in $v_{th}$, therefore assigning the higher-$v_{th}$ to the transistors in a circuit can be very useful in reducing the leakage current, and thereby reducing the Leakage Power [14]. But, the problem is that the higher-$v_{th}$ increases the equivalent ON-resistance for the transistors, and that in turn increases the Delay [14].

The propagation Delay through a transistor is generally denoted as,

$$T_{delay} = \frac{C_L V_{dd}}{K.(V_{dd} - V_{th})^\alpha} \quad (2)$$

Where, $K$ is a factor which depends on the gate size, as well as on the process. $\alpha$ takes any value between 1 and 2 depending on channel length [15].

Therefore, we can see that the reduction of the $v_{th}$ can be useful to improve the overall performance at low supply voltages [16]. But, as we reduce the $v_{th}$ of the transistor, leakage current starts playing a dominant role [13, 16]. Thus, maintaining the performance of the circuit as well as reducing the Leakage Power dissipation becomes a key challenge for designing any low-voltage, low power digital circuit.

Gate level DVT scheme is one of the most efficient techniques, for reducing the Leakage Power in a CMOS circuit. For a gate level DVT circuit, all the transistors within any particular gate may have the same $v_{th}$ which is either the low-$v_{th}$ or the high-$v_{th}$. Thus, the gates used in the logic circuit, can either be the low-$v_{th}$ gates or else, the high-$v_{th}$ gates [17].

## III. POWER GATING STRATEGIES

For the low leakage, high performance operation of any VLSI circuit, the Power Gating technique is treated as the most effective one which can substantially reduce the leakage current in standby mode. Now, considering the previously proposed circuit level approaches, the use of sleep transistors for Power Gating is found to be the most popular one [18-23]. When the circuit is in active mode these sleep transistors are 'ON'. But, for the standby mode of operation, these transistors get turned 'OFF', and that in turn disconnects the logic cells from the $V_{dd}$ (or, Ground) rail. Now, the major concern associated with the insertion of sleep transistors, is the degradation in performance [18]. It is found, for a specific placement technique, the amount of performance degradation of the circuit usually depends on the size of the sleep transistors [18]. In conventional Power Gating architecture, a 'header' and a 'footer' switch (which are basically the sleep transistors) used to be connected in series with the PUN (Pull-Up Network) and PDN (Pull-Down Network) of the logic gates respectively. As illustrated in Fig. 1, the virtual-$V_{dd}$ rail (virtual-Ground rail) could be disconnected from the actual $V_{dd}$ (Ground) by turning-off the 'header' ('footer') sleep-transistor; and thereby reducing the Leakage Power. But in active mode, these sleep transistors need to be turned 'ON', such that the logic circuit works fine as per its functionality. Now, instead of using both 'header' and 'footer' sleep transistors, the same Leakage Power reduction can be achieved by using any one of the two switches. Considering the perspective of area required, effective conductance etc., it is better to use NMOS sleep transistors as the footer switches [19, 20]. But, looking at the other-factors like noise on Power/Ground rails, Leakage etc., the implementation of PMOS header switches can also be very useful [22].

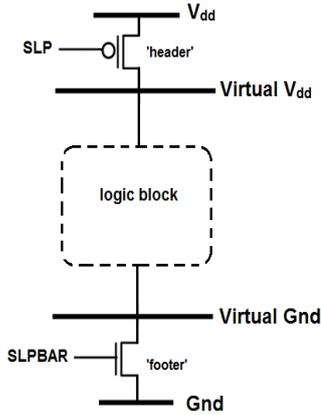

Figure 1. Conventional Power Gating architecture

As mentioned above, the effectiveness of the Power Gating technique actually depends on the proper sizing of the sleep transistors. For the larger sleep transistors, it can be seen that the performance degradation is lesser [18]. But simultaneously those larger transistors require larger area, and a significant amount of driving energy [23]. Whereas, the insertion of smaller sleep transistors may cause an increase in performance degradation, which is also not acceptable [18]. So, there is a trade-off in between the power consumption and the performance of the circuit. Several works have already been reported in literature to find out the feasible solution to mitigate the aforesaid problem [18-22]. DSTN, Clustering-based sleep transistor sizing algorithms etc. are the major among those. Besides, there is one other important technique, called Clock Gated Power Gating, where a clock signal is used to excite the sleep transistors. In case of the Clock Gated Power Gating scheme, the frequency of the clock signal is actually set depending up on the Delay among the intermediate results which are obtained from the different clusters of the circuit [11].

## IV. DESIGN OF THE 4-BIT BCD ADDER

As illustrated in Fig. 2, for designing the conventional 4 bit BCD adder we need two 4-bit full adders. These 4-bit full adders are basically Ripple Carry Adders (RCAs), where carry output of one full adder block is fed as the carry input for the next full adder. Besides that, for the purpose of forming the carry detection logic circuit, two AND gates and one OR gate have been used. Now, the full adder being the basic building block for designing the proposed 4-bit BCD adder, the proper selection of the 1-bit full adder cell becomes obvious. The design criteria of a full adder are actually multi-fold. Besides the transistor count which is one of the primary concerns, the two other important design criteria are the power consumption and the speed [24]. There are large numbers of full adder designs already available in literature, and those designs are relying upon different logic styles like, static CMOS, dynamic, transmission gate, or pass transistor logic [12], [16], [24-28]. However, considering the advantages like lesser Delay, elimination of the Short-Circuit Power component within the cell, fewer glitches at outputs, and importantly lesser power consumption, the circuit of the 16 transistor full adder which is reported in [28] has been adopted in this work. Fig. 3 shows the architecture of the 16 transistor 1-bit full adder cell where the actual design of [28] has been modified with the DVT scheme. In our proposed design of 4-bit BCD adder, we have actually used this modified 16 transistor 1-bit full adder cell for the purpose of implementing the two 4-bit full adder structures.

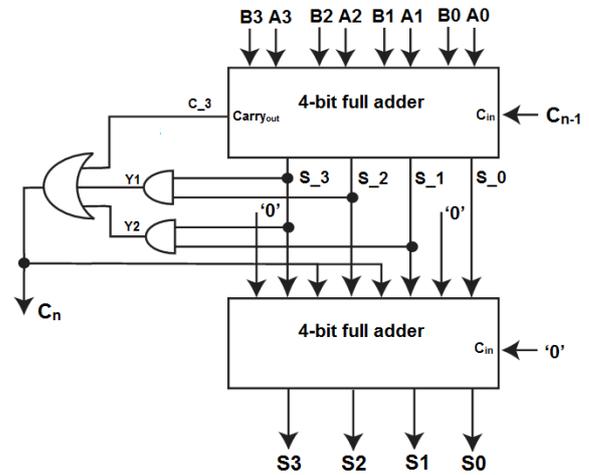

Figure 2. Conventional 4-bit BCD adder

Moreover, the concept of multiple channel length technique has also been utilized in the proposed architecture. For the conventional CMOS technology, the multiple channel length technique is known to be one of the popular means by which we can reduce the Leakage Power [29]. As per the technique, the channel length of the transistors used in a circuit can be increased, wherever it is needed to control the leakage current. On the other hand, wherever it is required to maintain the performance (specially, for the transistors in critical path), we need to increase the width of the transistors [29]. For the proposed design, the effective sizing of the sleep transistors and the transistors used in transmission gates has been done using this technique.

From the block level representation of the proposed 4-bit BCD adder (as shown in Fig. 4), it can be seen that the CLK1 and CLK2 are the two different clock signals which actually used for the purpose of Power Gating.

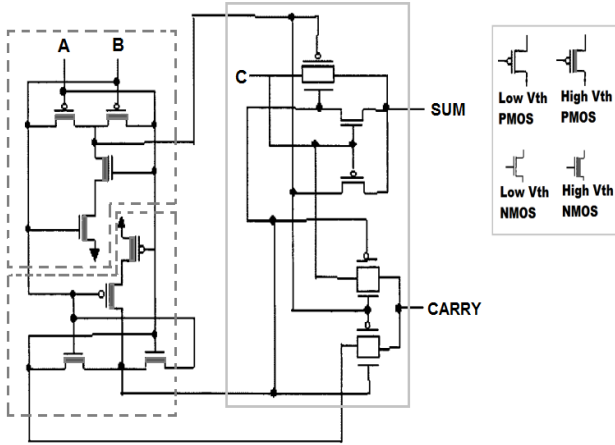

Figure 3. 16 transistor 1-bit full adder, modified with DVT scheme [28]

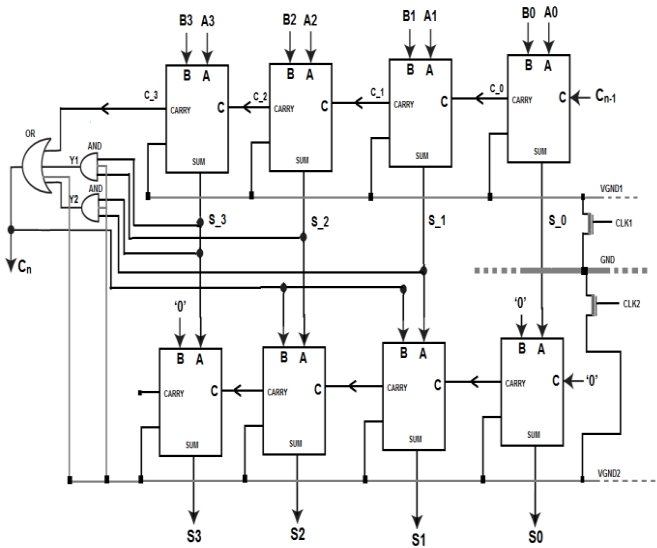

Figure 4. Block diagram of the proposed 4-bit BCD adder design

## V. RESULTS AND DISCUSSIONS

The analytical reasoning behind the modification of the conventional 4-bit BCD adder is verified here with circuit simulations. From Table I we are getting the details like power consumption, maximum Delay at output etc. for the proposed BCD adder along with the comparison of the proposed scheme with other state-of-the-art designs [30], [31]. Table II shows the comparison of the different performance parameters, for the proposed architecture with the conventional 4-bit BCD adder and the 4-bit BCD adder with DVT scheme (without Power Gating). The simulation results are obtained for an operating frequency of 200 MHz and a supply voltage of 1 Volt.

Table I. Different design styles for Low-Power BCD adders

|  | [30] (using CSLA) | [31] | This Work | Unit |
|---|---|---|---|---|
| Architecture | Flagged BCD adder | Unified adder/ subtractor | BCD adder designed using DVT, Clock Gated PowerGating |  |
| Technology | -- | 180 | 45 | nm |
| Operation | 4-bit addition | addition/ subtraction | 4-bit addition |  |
| Operating Frequency | -- | 1000 | 200 | MHz |
| Power | $2.55 \times 10^{-3}$ | $14.50 \times 10^{-3}$ | $1.384 \times 10^{-6}$ | Watt |
| Delay | $15.65 \times 10^{-9}$ | $3.46 \times 10^{-9}$ | $1.618 \times 10^{-10}$ | Second |

Table II. Comparison of Performances

|  | Conventional 4-bit BCD adder | 4-bit BCD adder with DVT (without Power Gating) | Proposed 4-bit BCD adder | Unit |
|---|---|---|---|---|
| Average Power | $3.722 \times 10^{-6}$ | $1.668 \times 10^{-6}$ | $1.384 \times 10^{-6}$ | Watt |
| Delay | $11.440 \times 10^{-11}$ | $19.229 \times 10^{-11}$ | $16.181 \times 10^{-11}$ | Second |
| (Average Power × Delay) | $42.588 \times 10^{-17}$ | $32.077 \times 10^{-17}$ | $22.394 \times 10^{-17}$ | Joule |

Fig. 5 shows the graphical representation of the details of Average Power consumption for the proposed 4-bit BCD adder, the conventional 4-bit BCD adder and the 4-bit BCD adder with DVT scheme (without Power Gating); for actually three different frequencies (50 MHz, 100 MHz, and 200 MHz). And as it is illustrated there, it can be observed that the proposed design performs extremely well for the higher values of frequencies, while consuming significantly lesser amount of Average Power. Furthermore, considering the products of Average Power and Delay, for the different frequency values, we can see that the proposed architecture provides the best case results (as shown in Fig.6).

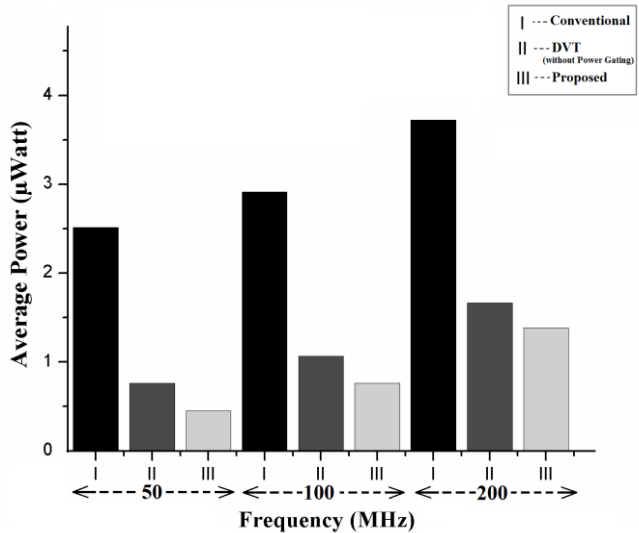

Figure 5. Average Power consumption details

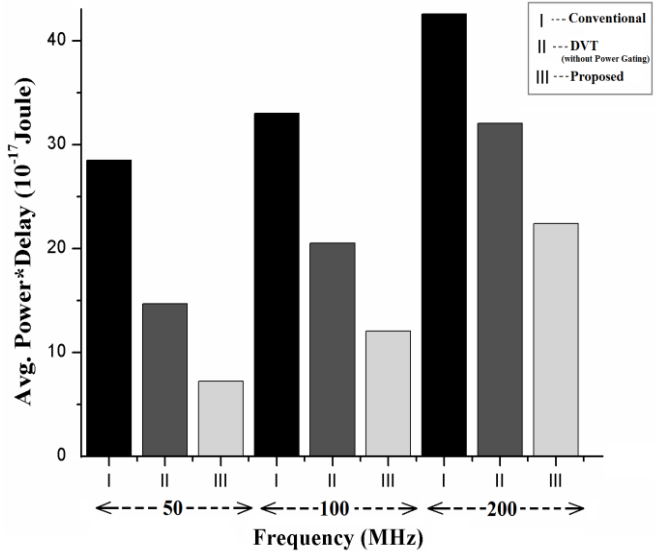

Figure 6. Products of Average Power and Delay, for different frequencies

## VI. CONCLUSION

Due to the limited battery life-time of the devices used in Low-Power applications, and the aggressive scaling of the transistor dimensions, the Leakage Power has become one of the major concerns while designing the digital circuits in VLSI. In this paper we have mainly focused to search for the means by which we can significantly reduce the power consumption of the circuit; and for that reason, we have modified the conventional 4-bit BCD adder architecture with Clock Gated Power Gating, and the DVT scheme.

Now, one of the possible drawbacks that the conventional BCD adder suffers from is the extra Delay at its output. And a large number of BCD adder designs have already been reported in literature [5], [10], [30] which can effectively be used to mitigate the problem. But, as we have mainly aimed to analyze the impact of the implementation of the proposed design scheme in improving the overall power performance, that is the why, here we have actually opted the conventional architecture.

From the results, as obtained from the different simulation runs on SPICE, it can be seen that a drastic reduction in Average Power consumption has become possible for the 4-bit BCD adder circuit, by using proposed design scheme. For an operating frequency of 200 MHz, the proposed architecture consumes 62.8 % less power compared to the conventional 4-bit BCD adder design.

ACKNOWLEDGMENT

Authors would like to thank SMDP-II project lab., IC Design & Fabrication Centre, Jadavpur University for getting the opportunity to carry out this work using SPICE Tools .